# High power, frequency agile comb spectroscopy in the mid-infrared enabled by a continuous-wave optical parametric oscillator

A. T. Heiniger[1*], M. J. Cich[1], D. A. Long[2*]

1TOPTICA Photonics, Inc., Pittsford, NY 14534, USA
2National Institute of Standards and Technology, Gaithersburg, MD 20899, USA
*Corresponding authors: A. T. Heiniger. Email: adam.heiniger@toptica-usa.com. D. A. Long. Email: david.long@nist.gov

**Abstract:** While mid-infrared optical frequency combs have been widely utilized in areas such as trace gas sensing, chemical kinetics, and combustion science, the relatively low power per comb tooth has limited acquisition times and sensitivities. We have developed a new approach in which an electro-optic frequency comb is utilized to pump a continuous-wave singly-resonant optical parametric oscillator in order to spectrally translate the comb into the mid-infrared. Through the use of electro-optic combs produced via chirped waveforms we have produced mid-infrared combs containing up to 2400 comb teeth. We show that a comb can be generated on the non-resonant idler when the pump modulation is non-synchronous, and we use these combs to perform high resolution spectroscopy on methane. In addition, we describe the underlying theory of this method and demonstrate that phase matching should allow for combs as broad as several THz to be spectrally translated to the mid-infrared. The high power and mutual coherence as well as the relatively low complexity of this approach should allow for broad application in areas such as chemical dynamics, quantum information, and photochemistry.

## 1. Introduction

Molecular spectroscopy at mid-infrared wavelengths has been widely utilized for gas sensing [1], chemical dynamics [2], quantum information [3], and astrochemistry [4] due to the large absorption cross-sections present in this region. Direct optical frequency comb spectroscopy (DCS) can offer advantages in spectral bandwidth, resolution, and acquisition time over other spectroscopic techniques, but it is difficult to simultaneously realize these capabilities in the mid-infrared. In particular, while broadband, high-resolution DCS has been demonstrated in the mid-infrared, the low power per comb tooth has largely precluded fast acquisition times and high sensitivities [5], [6].

Achieving higher power per comb tooth in the mid-infrared has been an ongoing challenge. Mode-locked quantum cascade lasers (QCLs) can provide milliwatts of power per tooth, but gigahertz tooth spacings limit the spectral resolution [7]. High-resolution mid-infrared comb approaches have traditionally offered tens to hundreds of nanowatts of power per comb tooth [8], [9], with at most 100 μW per comb tooth generated through the use of a 57 W pump laser [8]. Recently, we have developed a new approach for mid-infrared comb generation in which a continuous-wave singly-resonant optical parametric oscillator (CWSRO) is used for frequency conversion of near-infrared electro-optic frequency combs [9]. This approach allows for high power per tooth (approximately 80 mW), and correspondingly high acquisition rates (up to 50 MHz) while also offering broad wavelength tunability.

These results indicate that this method may fulfill the promise of fast, sensitive absorption spectroscopy in the mid-infrared. In the present study, we describe the theory governing the frequency conversion of the modulated pump and derive the conditions under which the pump comb is replicated on the non-resonant, mid-infrared CWSRO idler. Our initial DCS demonstration [9] utilized over-driven electro-optic phase modulators (EOMs) to produce roughly twelve comb teeth in each frequency comb with a comb tooth spacing of 2.55 GHz. Here we pump the CWSRO with a single electro-optic frequency comb combined with a local oscillator. To generate this comb, a near-infrared EOM is driven with chirped waveforms to produce an optical frequency comb with up to 2400 individual comb teeth of similar amplitude. This flat, high-resolution comb allows us to experimentally confirm the derived theory for replication of the pump comb on the idler, and results in a mid-infrared comb that is widely tunable between 2.19 µm and 4.00 µm with Watt-level power and approximately 450 µW per comb tooth in a relatively simple setup. We have then utilized the comb spectrometer to perform high resolution molecular spectroscopy in the critical mid-infrared spectral region.

## 2. Theory

In this section we derive the conditions under which the spectrum of a pump laser is replicated on the non-resonant idler of a CWSRO. In particular we show that the pump spectrum should contain no spectral components which are separated by the signal cavity's free spectral range (FSR), and that the phase matching bandwidth of the CWSRO provides the ultimate limit on the bandwidth of the idler spectrum.

We follow the derivation of Yariv for three waves interacting in a nonlinear medium [10]. There, differential equations are derived assuming that the three waves have no time dependence except oscillation at their carrier frequencies. Here we allow for variation in time at rates much slower than the pump carrier frequency. The real electric field $\mathcal{E}$ for pump, signal, and idler waves $j = (p, s, i)$ is written in terms of a complex field $E$ which varies slowly in time and space:

$$\mathcal{E}_j(z,t) = \tfrac{1}{2} E_j(z,t) e^{i\omega_j t} e^{-ik_j z} + c.c. \tag{1}$$

where $\omega_j$ are the center angular frequencies of the three waves and $k_j$ are their wavevectors.

This expression for the electric field represents a plane wave, and we model the field as such for simplicity. In practice, the pump, signal, and idler in our CWSRO have spatial modes defined by the pump laser and optical cavity. However, this plane wave approximation has adequately modeled the behavior of similar CWSROs [11], [12], [13], and the transfer of modulation to the idler that we explore here is independent of the spatial mode.

We employ Type 0 phase matching in periodically-poled lithium niobate (PPLN), so that all electric field components have the same polarization. The equations which govern propagation of the pump, signal, and idler in the nonlinear crystal are then given by

$$\frac{dE_p}{dz} = -\frac{n_p}{c}\frac{\partial E_p}{\partial t} - 2\frac{d_{eff}}{n_p c}\frac{\partial (E_s E_i)}{\partial t} e^{i\Delta k z} - i\frac{\omega_p d_{eff}}{n_p c} E_s E_i e^{i\Delta k z} \tag{2}$$

$$\frac{dE_s}{dz} = -\frac{n_s}{c}\frac{\partial E_s}{\partial t} - 2\frac{d_{eff}}{n_s c}\frac{\partial (E_p E_i^*)}{\partial t} e^{-i\Delta k z} - i\frac{\omega_s d_{eff}}{n_s c} E_p E_i^* e^{-i\Delta k z} \tag{3}$$

$$\frac{dE_i}{dz} = -\frac{n_i}{c}\frac{\partial E_i}{\partial t} - 2\frac{d_{eff}}{n_i c}\frac{\partial (E_p E_s^*)}{\partial t} e^{-i\Delta k z} - i\frac{\omega_i d_{eff}}{n_i c} E_p E_s^* e^{-i\Delta k z} \tag{4}$$

where $n_j$ are the nonlinear crystal refractive indices at $\omega_j$, and c is the speed of light in vacuum. The wavevector mismatch is $\Delta k = k_p - k_s - k_i - K_g$, where $K_g = 2\pi/\Lambda_g$ is the wavevector of the quasi-phase matching (QPM) grating of period $\Lambda_g$, and $d_{eff}$ is the effective nonlinear coefficient for ideal first-order QPM. In PPLN $d_{eff} = 17$ pm/V [14].

We are concerned with the evolution of the pump, signal, and idler spectra, so we take the Fourier transform of these equations with respect to angular frequency detuning $\Omega$ to obtain

$$\frac{d\tilde{E}_p}{dz}(z,\Omega) + \frac{i\Omega n_p}{c}\tilde{E}_p(z,\Omega) = -i\frac{(\omega_p+2\Omega)d_{eff}}{n_p c}e^{i\Delta k z}\int_{-\infty}^{\infty}d\Omega'\tilde{E}_s(z,\Omega')\tilde{E}_i(z,\Omega-\Omega') \qquad (5)$$

$$\frac{d\tilde{E}_s}{dz}(z,\Omega) + \frac{i\Omega n_s}{c}\tilde{E}_s(z,\Omega) = -i\frac{(\omega_s+2\Omega)d_{eff}}{n_s c}e^{-i\Delta k z}\int_{-\infty}^{\infty}d\Omega'\tilde{E}_p(z,\Omega')\tilde{E}_i^*(z,\Omega-\Omega') \qquad (6)$$

$$\frac{d\tilde{E}_i}{dz}(z,\Omega) + \frac{i\Omega n_i}{c}\tilde{E}_i(z,\Omega) = -i\frac{(\omega_i+2\Omega)d_{eff}}{n_i c}e^{-i\Delta k z}\int_{-\infty}^{\infty}d\Omega'\tilde{E}_p(z,\Omega')\tilde{E}_s^*(z,\Omega-\Omega'). \qquad (7)$$

As each wave in the nonlinear crystal propagates it generates new spectral components from the convolution of the spectra of the other two waves.

The nonlinear crystal is located inside an optical cavity which is resonant only for the signal wave. The optical cavity supports signal frequencies which correspond to longitudinal modes, and all other signal frequencies will interfere destructively. If the signal oscillates on only a single longitudinal cavity mode, then its spectrum is practically a delta function with respect to the pump spectrum and Eqn. 7 for the idler becomes

$$\frac{d\tilde{E}_i}{dz}(z,\Omega) + \frac{i\Omega n_i}{c}\tilde{E}_i(z,\Omega) = -i\frac{(\omega_i+2\Omega)d_{eff}}{n_i c}e^{-i\Delta k z}\tilde{E}_p(z,\Omega). \qquad (8)$$

Therefore, if the signal is single mode, then a spectral component detuned $\Omega$ from the pump carrier will be generated on the idler detuned $\Omega$ from the idler carrier. Thus, the frequency spacing of the idler spectral components will match that of the pump. The relative amplitudes of these spectral components on the idler nearly match the pump. The $\omega_i + 2\Omega$ factor on the right-hand side of Eqn. 8 will slightly skew the idler spectrum. As described below, the maximum detuning considered here is on the order of 1 THz, which is much smaller than the 75 THz to 135 THz idler carrier frequency. Thus, if the signal is single mode, then the idler spectrum will replicate the pump spectrum except with a slope in amplitude of approximately 1%.

We note that there are modulation conditions which can cause the signal to oscillate on multiple modes. An etalon in the CWSRO cavity is used to force single longitudinal mode operation in the case of an unmodulated pump. The etalon has a bandpass full-width at half maximum (FWHM) of approximately 30 GHz and the cavity has an FSR near 530 MHz. One cavity mode under the etalon bandpass has the highest gain and dominates when the pump is unmodulated, but approximately 60 modes have significant gain. When the pump is modulated, then cavity modes which neighbor the highest-gain mode also can oscillate.

An extreme case of pump modulation coupling to the resonant wave is a synchronously pumped OPO [17], [18]. There, the pump and OPO cavity are stabilized so that the comb mode spacing exactly matches the cavity FSR. From Eqns. 5-7, a comb spectrum on the pump and signal leads to cascaded generation of comb teeth on the pump, signal, and idler.

To ensure that only a single cavity mode oscillates we avoid synchronous modulation. The signal oscillates on a center cavity mode at $\omega_s$, and in general it can carry modulation and have spectrum $\tilde{E}_s(z,\Omega)$. However, the signal can only support modulation at frequencies which correspond to longitudinal modes, where the detuning $\Omega$ is equal to multiples of the cavity FSR. If the pump spectrum $\tilde{E}_p(z,\Omega)$ does not have any spectral components which are separated by multiples of the cavity FSR, then pump modulation cannot couple to the signal cavity.

Thus, the first condition for successful replication of the pump spectrum on the idler of a CWSRO is that the pump spectrum does not contain spectral components which are separated by the cavity FSR. The second condition relates to the bandwidth of the pump comb. The PPLN poling period can be chosen so that $\Delta k = 0$ at the center frequencies of the pump, signal, and idler. However, the second term in Eqn. 3 provides a first-order correction to the phase accumulation in each of the three waves due to dispersion. This will result in non-zero Δ$k$ for spectral components which are detuned from the center frequencies. This phase mismatch results in decreased gain, and thus decreased power for idler comb teeth. The full-width at half-maximum (FWHM) of the phase matching gain curve for the CWSRO used here was calculated by SNLO [15] and is plotted as a function of idler wavelength in Fig. 1. It shows that the widest FWHM of the idler spectrum is several terahertz when the idler wavelength is near the degeneracy point and decreases to approximately 100 GHz far from degeneracy.

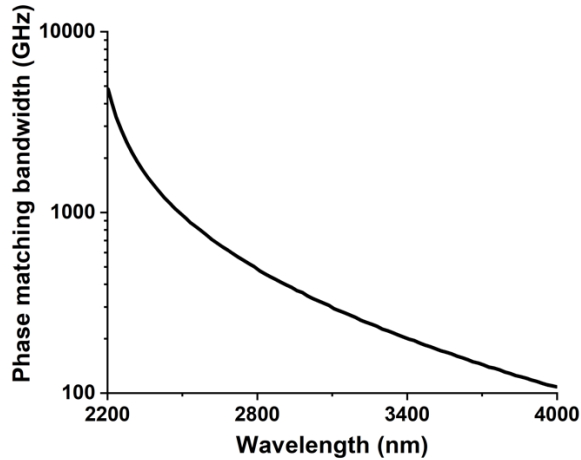

Figure 1. Full-width at half-maximum phase matching bandwidth of the idler of the continuous-wave singly resonant optical parametric oscillator pumped at 1064 nm.

We note that the CWSRO could be optimized for broader phase matching. A similarly designed CWSRO with a broadband incoherent pump utilized non-collinear phase matching via tight focusing to generate an idler that was 5.7 THz wide (FWHM) at 3.4 µm [16].

## 3. Experiment

A schematic of the present mid-infrared optical frequency comb spectrometer can be found in Figure 2a. Light from a continuous-wave external cavity diode laser with a wavelength near 1064 nm had its fiber-coupled output split into two paths: a probe path and a local oscillator (LO) path in a self-heterodyne configuration [17], [18]. An electro-optic frequency comb is produced on the probe path by driving an EOM with the output of a chip-scale direct digital synthesizer (DDS) [19]. The DDS generates periodic frequency chirps as shown in Fig. 2b and accepts sweep and timing information from a programmable microcontroller. This periodic phase modulation results in a frequency comb where the comb tooth spacing is equal to the chirp repetition rate and the comb bandwidth is twice the bandwidth of the chirp. Importantly, this approach allows for agile, ultraflat frequency combs to be generated over a wide range of comb tooth spacings (100's of Hz to GHz) [19].

An acousto-optic modulator (AOM) placed on the LO path provided a 54.2 MHz shift to ensure that the positive and negative order comb teeth occur at unique radiofrequencies once combined on a photodiode. A representative near-infrared self-heterodyne comb spectrum can be found in Fig. 2c. 240 comb teeth which are spaced by 10 MHz can be observed as well as the carrier tone which occurs at the AOM frequency.

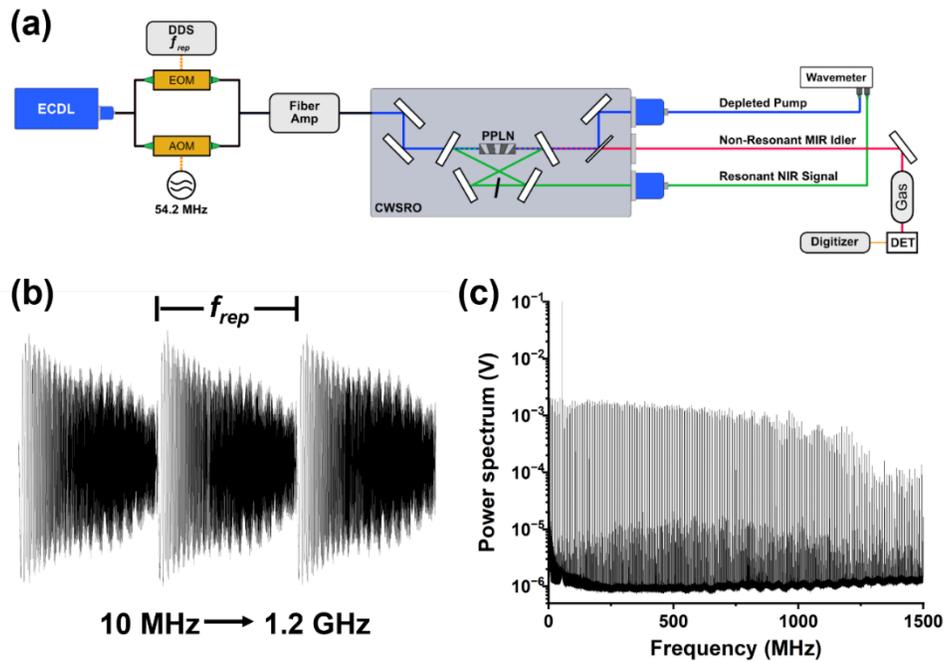

Figure 2. (a) Schematic of the continuous-wave optical-parametric-oscillator-based optical frequency comb spectrometer. The output of an external-cavity diode laser (ECDL) is split into two paths, one containing an electro-optic phase modulator (EOM) and the other an acousto-optic modulator (AOM) which serve as probe and local oscillator (LO) paths, respectively. The EOM is driven by a periodic chirp to generate an electro-optic frequency comb. The chirp repetition rate ($f_{rep}$) and frequency range are controlled by the output of a direct digital synthesis (DDS) chip (see panel (b)). The LO path passes through an AOM to produce a frequency shift of 54.2 MHz. The probe and LO paths are then recombined and amplified with a fiber amplifier before being injected into the continuous-wave optical parametric oscillator (CWSRO) containing a periodically poled lithium niobate (PPLN) crystal. The resulting mid-infrared (MIR) idler frequency comb was attenuated and then passed through a gas cell containing 48 Pa of methane before being detected on a photodetector (DET). A representative near-infrared (NIR) self-heterodyne spectrum can be seen in panel (c).

The combined probe and LO paths were then amplified by an ytterbium fiber amplifier which increased the seed power from 5 mW to 10 W. This amplified pump was then injected into the CWSRO which generates tunable mid-infrared radiation by down-converting near-infrared pump photons into near-infrared signal and mid-infrared idler photons through the use of a PPLN crystal. The CWSRO used here has been previously described in detail in Ref. [20].

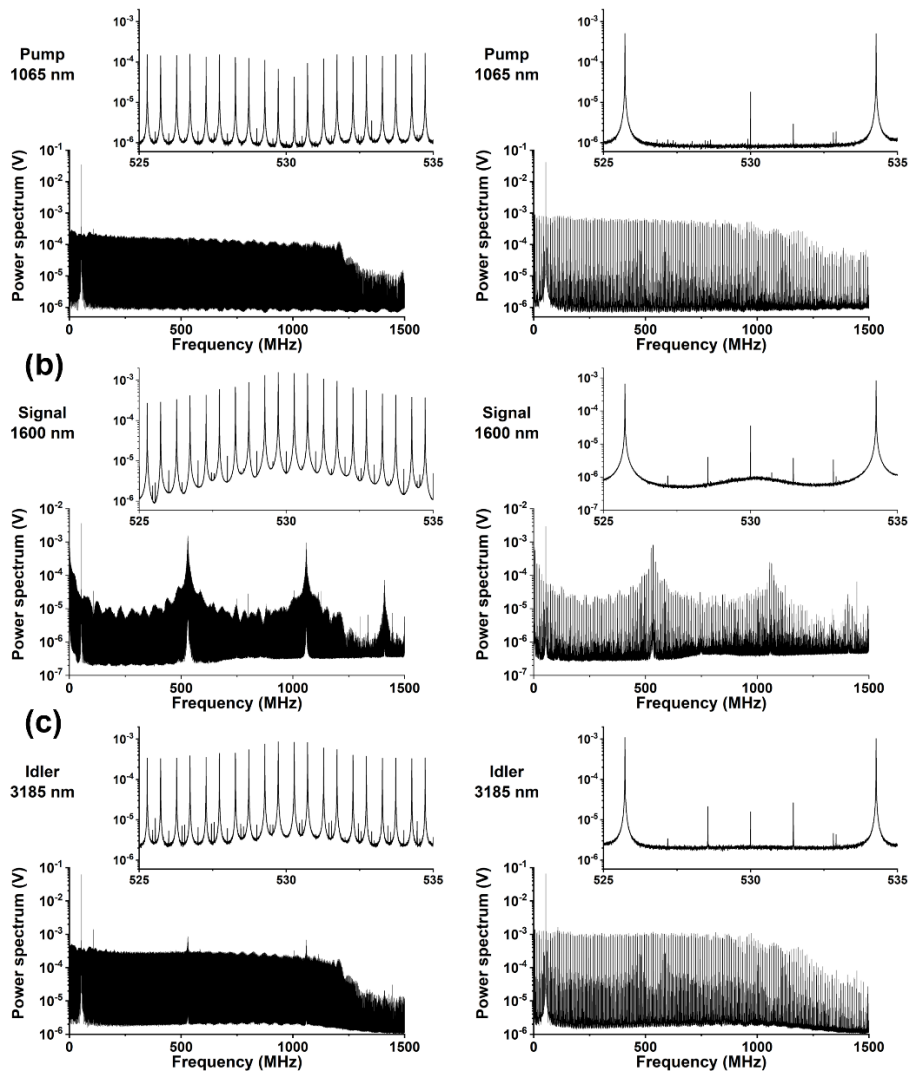

Figure 3. Radiofrequency spectra of the pump (a), signal (b), and idler (c) when driven by near-infrared electro-optic combs with 1 MHz (left panels) and 10 MHz repetition rate (right panels). The insets show magnified versions of the radiofrequency spectra which are centered near the OPO cavity's free spectral range. Each of the shown power spectra is the average of one hundred power spectra each of which were acquired in 0.5 ms.

The poling period of the PPLN varies along the crystal height in a fan-out structure. Vertical translation of the crystal position relative to the pump beam changes the phase matching conditions, which widely tunes the signal (1450 nm to 2070 nm) and idler (2190 nm to 4000 nm). Rotation of the etalon and continuous tuning of the seed laser finely tunes the idler to a target mid-infrared wavelength. The depleted pump and signal beam were sent to a wavemeter for a continuous wavelength measurement. Measurement of the pump and signal wavelengths with 150 MHz accuracy allowed calculation of the idler wavelength with 210 MHz accuracy. The CWSRO had a nominal output power of 2 W in the depleted pump, 3 W in the signal, and 2 W in the idler.

## 4. Results and Discussion

The three outputs of the CWSRO (depleted pump, signal, and idler) were recorded on fast photodiodes using a digitizer operating at 3 gigasamples/s. Fast Fourier transforms (FFTs) of these interferograms can be found in Figure 3 for both 1 MHz and 10 MHz comb spacings (left and right panels, respectively). The depleted pump spectrum contains both the carrier tone at 54.2 MHz as well as the 2400-MHz-wide comb which was initially produced in the near-infrared prior to amplification (i.e., Fig. 2c). The shown pump combs contained 2400 and 240 individual comb teeth for the 1 MHz and 10 MHz comb spacing cases, respectively. In order to further visualize the flatness and extent of the optical frequency combs we extracted the FFT magnitudes at the known comb tooth frequencies and normalized them against the carrier tone for the 1 MHz pump, signal, and idler combs (see Fig. 4). The shown idler combs had powers per comb tooth of approximately 40 µW and 450 µW for the 1 MHz comb and 10 MHz combs, respectively, and are therefore well suited for high signal-to-noise molecular spectroscopy.

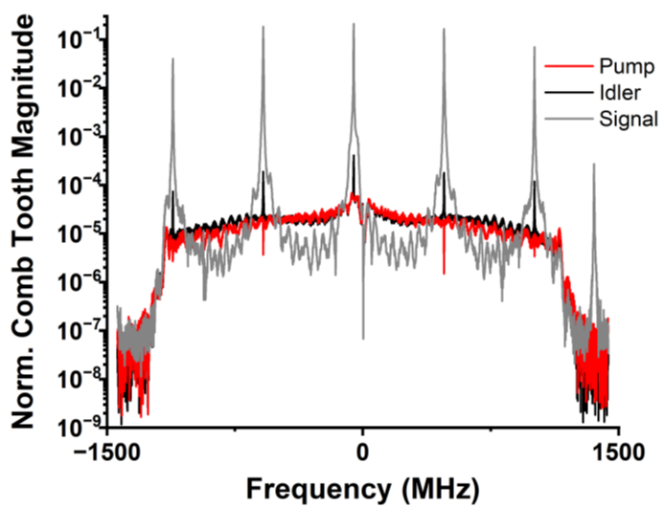

Figure 4. Comb tooth magnitudes normalized against the carrier tone for the pump, idler, and signal outputs of the CWSRO for a 1 MHz spaced comb. The shown spectra are the average of ten spectra each of which was acquired in 1 ms.

As previously described, complete replication of the pump comb on the idler requires that the pump comb does not contain spectral components spaced by multiples of the cavity FSR. The EO comb carrier and the 54.2 MHz-detuned LO are the dominant components of the pump, so separations from these components at multiples of the cavity FSR are the dominant contributors to multimode oscillation of the signal. While the 10 MHz spaced pump comb is replicated with minimal distortion on the idler, the 1 MHz spaced comb contains teeth which are separated from the comb carrier and LO by multiples of the cavity FSR. As a result, we see significant depletion of the pump and enhancement of the signal and idler at multiples of 530 MHz (see Fig. 3b and 4). Weaker features also can be observed which arise at frequencies which are separated from the LO frequency by the cavity FSR (e.g., 476 and 584 MHz). The signal should only have gain at harmonics of the cavity FSR, so the observation of comb teeth on the signal at frequencies between multiples of the cavity FSR was unexpected. We believe these weak comb teeth are present as there can be some transfer of phase to the signal in a single pass through the crystal.

As an initial spectroscopic demonstration, we passed the idler beam through a 40 cm long cell containing 48 Pa of methane. Normalization of the resulting comb spectrum was performed via a background spectrum recorded when the cell was removed. The resulting transmission spectrum is plotted in Fig. 5 and compared to a HITRAN 2020 [21] fit in which only the center wavelength and background level were floated to account for uncertainty associated with the wavelength meter and optical cell coupling losses. This spectrum contains 2400 individual comb teeth and was recorded in only 0.5 s.

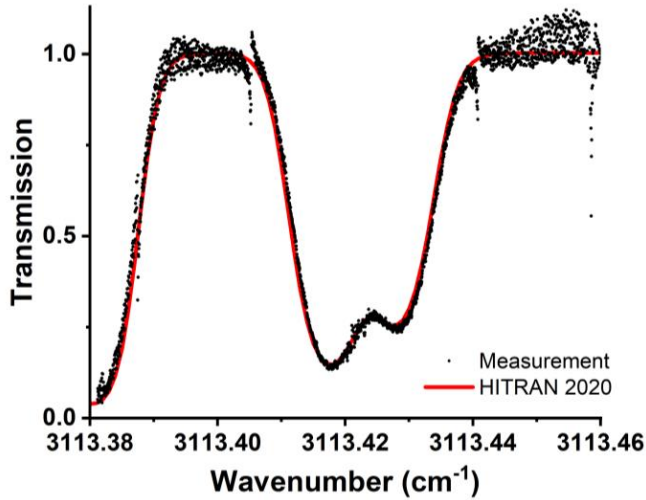

Figure 5. Transmission spectrum (black points) of a gas cell containing 48 Pa of methane as well as a spectral fit (red line) using HITRAN 2020 [21] parameters. Only the center wavelength and background level were adjusted to account for uncertainty in the wavelength meter and coupling losses of the optical cell. The shown spectrum contains 2400 individual comb teeth spaced by 1 MHz and was the average of 500 spectra each of which was acquired in 1 ms. As predicted by the theory, small distortions in the spectrum are observed for comb teeth occurring at multiples of the CWSRO cavity's free spectral range from the carrier wavelength.

The use of the 1 MHz comb for this demonstration allows us to see the impact of the slight idler comb perturbations when used for spectroscopy. Distortions are clearly present on the recorded spectrum at multiples of the cavity FSR, but the features viewed here are much broader than these distortions and an accurate spectrum can be measured. We note that these distortions can be removed by using a wide spaced comb (e.g., 10 MHz) which does not contain comb components at the cavity FSR.

We believe that the present technique holds numerous advantages in comparison to other mid-infrared comb approaches such as QCL combs [7], difference frequency generation (DFG) combs [8], [22], [23], and synchronously pumped femtosecond OPO-based combs (SySRO combs) [24], [25]. The present approach offers far higher tunability and agility with respect to repetition rate than QCL or SySRO combs where the repetition rate is essentially fixed and limited by either the QCL cavity length (generally near 10 GHz) [7] or the SySRO cavity length (generally a few hundred MHz). The repetition rates used in the present work are determined by the driving frequencies of an EOM, which are frequency agile and digitally controlled. Thus, a single EOM-comb CWSRO system can be used for applications requiring spectral resolution ranging from well less than 1 MHz to more than 1 GHz. In addition, since the local oscillator (or a second comb) can be generated from the same near-infrared laser and simultaneously translated into the mid-infrared by the same CWSRO, there is no need for complicated phase locking, phase correction, or a second comb source, in contrast to these other methods. Finally, in comparison to DFG combs the present method offers far higher optical powers for a given pump power [8], [22], [23].

One area of future work will be extending the bandwidth of the pump combs to reach the bandwidth limits imposed by the phase matching. Broader pump combs could be generated via cascaded modulators or non-linear spectral broadening. As shown earlier, the phase matching condition of the CWSRO is very broad and is expected to accommodate combs as wide as several THz, allowing for wideband multiplexed spectroscopy.

The combination of high resolution, high power, and mutual coherence provided by CWSRO EOM combs is ideally suited to applications such as sub-Doppler spectroscopy, where narrow features must be located within a broad spectral region. In addition, we see strong applications for this approach in areas such as chemical kinetics, optical metrology, communications, and quantum sensing where the flexibility, agility, and high optical power are expected to be transformative.

**Acknowledgements.** We thank D. B. Foote and J. T. Hodges for helpful discussions. Certain equipment, instruments, software, or materials are identified in this paper in order to specify the experimental procedure adequately. Such identification is not intended to imply recommendation or endorsement of any product or service by NIST, nor is it intended to imply that the materials or equipment identified are necessarily the best available for the purpose.

**Disclosures.** ATH: TOPTICA Photonics, Inc. (E,P), MJC: TOPTICA Photonics, Inc. (E,P).

**Data availability.** The data underlying this paper will be publicly available at a data.nist.gov.